\documentclass[a4paper,fleqn]{article}
\usepackage{a4,graphicx,cite}

\makeatletter
\renewcommand{\@biblabel}[1]{#1.}
\makeatother

\begin{document}

\title{\bf Self-consistency and vertex corrections beyond the {\boldmath $GW$}
  approximation}
\author{Arno Schindlmayr\thanks{Correspondence: Dr.\ Arno Schindlmayr,
  Fritz-Haber-Institut der Max-Planck-Gesellschaft, Faradayweg 4--6, 14195
  Berlin-Dahlem, Germany. E-mail: schindlmayr@fhi-berlin.mpg.de}}
\date{Fritz-Haber-Institut der Max-Planck-Gesellschaft, Faradayweg 4--6,
  14195 Berlin-Dahlem, Germany}
\maketitle

\begin{abstract}
The good performance of the $GW$ approximation for band-structure calculations
in solids was long taken as a sign that the sum of self-energy diagrams is
converged and that all omitted terms are small. However, with modern
computational resources it has now become possible to evaluate
self-consistency and vertex corrections explicitly, and the numerical results
show that they are, in general, not individually negligible. In this review
the available data is examined, and the implications for practical
calculations and the theoretical foundation of the $GW$ approximation are
discussed.
\end{abstract}

\section{Introduction}

Many-body perturbation theory \cite{Mahan1990} represents a powerful method
for studying the properties of interacting electron systems from first
principles that goes beyond the limits of local mean-field approaches. It is
based on the Green function, which can be interpreted as a propagator that
describes the evolution of an additional electron or hole injected into the
system and interacting with its environment via the Coulomb potential. In this
way the Green function can be directly related to experimental photoemission
spectra in solid state physics, and macroscopic observables like the total
energy are obtained through an integral with the respective
quantum-mechanical operator. Furthermore, unlike the Kohn-Sham eigenvalues in
density-functional theory \cite{Hohenberg1964,Kohn1965}, the resonances of the
Green function have a well defined physical meaning as electron addition or
removal energies and correspond to the proper quasiparticle band
structure. They can be calculated from a modified single-particle
Schr\"odinger equation, in which exchange and correlation effects are
rigorously described by the so-called self-energy operator, which takes the
form of a nonlocal, energy-dependent potential \cite{Hedin1969}.

A practical way of calculating the self-energy is given by a perturbation
expansion in terms of the Coulomb potential and the Green function of the
corresponding noninteracting system \cite{Mahan1990}. This expansion is most
conveniently written in the language of Feynman diagrams \cite{Feynman1949},
which may be related to distinct scattering mechanisms. As it is not possible
to sum the complete infinite perturbation series, physical intuition can serve
to identify the most important contributions. Of course, the selection of
diagrams depends on the nature of the problem. In simple metals and
semiconductors correlation is predominantly long-range, because electrons
interact with their environment through polarization of the surrounding medium
and thus avoid proximity. This effect is well described by Hedin's $GW$
approximation \cite{Hedin1965}, which includes dynamic polarization in the
random-phase approximation.

The performance of the $GW$ approximation has been discussed in several recent
reviews \cite{Aryasetiawan1998,Aulbur2000a}. Most importantly, it corrects the
systematic underestimation of semiconductor band gaps in Kohn-Sham
density-functional theory, giving very good agreement with experimental data
\cite{Hybertsen1985,Godby1986}. Another more subtle effect relates to the
narrowing of the occupied band width in the alkali metals, which is also
described accurately \cite{Northrup1987,Surh1988}. Because of this success it
was long believed that the $GW$ approximation indeed captures all relevant
self-energy terms for these materials and that the excluded diagrams have
negligible weight. However, recent numerical studies that explicitly evaluated
self-consistency and vertex corrections, the principal omissions in the $GW$
approximation, have cast doubts on this assumption and suggest that the high
quantitative accuracy instead stems from a cancellation of errors. A better
understanding of these effects would not only strengthen the theoretical
foundation of the $GW$ approximation, but might also show the way to further
systematic improvements. For this reason I review recent progress in this
field below and assess the available data.

This paper is organized as follows. In Sec.\ \ref{Sec:GW} the $GW$
approximation is introduced. Self-consistency and vertex corrections are
discussed in Secs.\ \ref{Sec:self-consistency} and \ref{Sec:vertex},
respectively. Finally, the main points are summarized in Sec.\
\ref{Sec:summary}. Atomic units are used throughout.

\section{The {\boldmath $GW$} approximation}\label{Sec:GW}

Many-body perturbation theory is based on the Green function, which is defined
as the expectation value of the time-ordered operator product \cite{Mahan1990}
\begin{equation}
G(1,2) = -i \langle \Psi | T[ \hat{\psi}(1) \hat{\psi}^\dagger(2) ] | \Psi
\rangle \;.
\end{equation}
The short-hand notation $(1) \equiv ({\bf r}_1,\sigma_1,t_1)$ indicates a set
of spatial, spin and temporal coordinates, $| \Psi \rangle$ denotes the
normalized ground-state wavefunction in the Heisenberg picture, and $T$ is
Wick's time-ordering operator that rearranges the subsequent symbols in
ascending order from right to left. Besides, $\hat{\psi}^\dagger(2)$ and
$\hat{\psi}(1)$ represent the electron creation and annihilation operator in
the Heisenberg picture, respectively. In the absence of a time-dependent
external potential, the Green function only depends on the difference $t_1 -
t_2$ and can be mapped to frequency space through a one-dimensional Fourier
transform. In the case of noninteracting systems, where the wavefunction is a
single Slater determinant, this yields the expression
\begin{equation}
G_0({\bf r}_1,{\bf r}_2;\omega) = \sum_{n,{\bf k}} \frac{\varphi_{n{\bf
  k}}({\bf r}_1) \varphi_{n{\bf k}}^*({\bf r}_2)}{\omega - \epsilon_{n{\bf k}}
  + i \,\mbox{sgn}(\epsilon_{n{\bf k}} - \mu) \eta}
\end{equation}
in terms of the solutions
\begin{equation}
\left( -\frac{1}{2} \nabla^2 + V_s({\bf r}) \right) \varphi_{n{\bf k}}({\bf
  r}) = \epsilon_{n{\bf k}} \varphi_{n{\bf k}}({\bf r})
\end{equation}
of the single-particle Schr\"odinger equation. Here $\mu$ denotes the chemical
potential that separates occupied from unoccupied states and $\eta$ is a
positive infinitesimal. Spin variables have been suppressed, because in the
absence of magnetization $G_0$ is symmetric and diagonal in $\sigma$. In the
following it is assumed that the single-particle potential $V_s$ already
includes the Hartree potential.

The Green function of an interacting electron system is related to the
propagator of the corresponding Hartree system through Dyson's equation
\cite{Dyson1949}
\begin{equation}\label{Eq:Dyson}
G(1,2) = G_0(1,2) + \int\! G_0(1,3) \Sigma(3,4) G(4,2) \,d(34) \;.
\end{equation}
The self-energy operator $\Sigma$ rigorously describes all exchange and
correlation effects. It can be expanded in a perturbation series comprising
all connected and topologically distinct diagrams constructed from $G_0$ and
the Coulomb potential $v$. As an alternative, Hedin \cite{Hedin1965} derived a
set of exact integral equations that define $\Sigma$ as a functional of
$G$. This constitutes a closure relation, which, in combination with Dyson's
equation, allows a self-consistent algebraic determination of the Green
function. As intermediate quantities, Hedin's equations
\begin{eqnarray}
\Sigma(1,2) &=& i \int\! G(1,3) W(1^+,4) \Gamma(3,2;4) \,d(34) \;,
\label{Eq:self-energy} \\
W(1,2) &=& v(1,2) + \int\! W(1,3) P(3,4) v(4,2) \,d(34) \;, \\
P(1,2) &=& -i \int\! G(2,3) G(4,2) \Gamma(3,4;1) \,d(34) \;, \\
\Gamma(1,2;3) &=& \delta(1,2) \delta(1,3)  + \int\! \frac{\delta
  \Sigma(1,2)}{\delta G(4,5)} G(4,6) G(7,5) \Gamma(6,7;3) \,d(4567) \label{Eq:vertex}
\end{eqnarray}
employ the screened Coulomb interaction $W$, the polarizability $P$ and the
vertex function $\Gamma$. The notation $(1^+)$ indicates that a positive
infinitesimal is added to the time variable.

In principle, this set of equations could be solved by iteration from a
suitable starting point, such as $G_0$, until self-consistency is
reached. However, the occurrence of a functional derivative in Eq.\
(\ref{Eq:vertex}) prevents an automatic numerical solution, because it changes
the mathematical expression for the integrand in each loop. In practice this
means that the starting point must be chosen so close to the expected solution
that self-consistency is reached after a very small number of iterations, for
which the functional derivative can be evaluated analytically. For solids the
Hartree Green function seems a good enough starting point, and solving Hedin's
equations with $\Sigma = 0$ then produces the so-called $GW$ approximation
\cite{Hedin1965}
\begin{equation}\label{Eq:GW}
\Sigma_{GW}(1,2) = i G(1,2) W(1^+,2)
\end{equation}
after one iteration. The screened interaction enters in the time-dependent
Hartree or random-phase approximation, in which the polarization propagator is
given by
\begin{equation}\label{Eq:RPA}
P_{\rm RPA}(1,2) = -i G(1,2) G(2,1) \;.
\end{equation}
The $GW$ approximation can be regarded as a generalization of the nonlocal
Fock or exchange potential $\Sigma_{\rm x}(1,2) = i G(1,2) v(1^+,2)$ with
dynamically screened exchange. In addition to Pauli's principle it includes
polarization effects and therefore describes correlation between electrons
with parallel spin as well as electrons with opposite spin.

\begin{figure}
\centerline{\includegraphics[width=6.5cm]{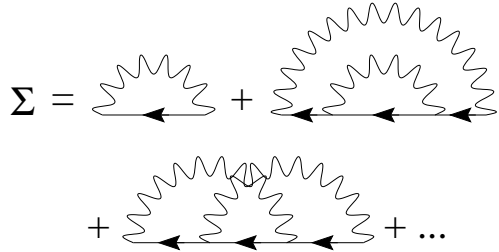}}
\renewcommand{\figurename}{\bf Figure}
\caption{\rm Diagrammatic expansion of the self-energy. Arrows represent
  noninteracting Green functions, the screened interaction is indicated by a
  wiggly line.}\label{Fig:diagrams}
\end{figure}

To be consistent with the iterative solution of Hedin's equations, the $GW$
approximation should be evaluated with the Hartree Green function $G_0$,
although in practice the corresponding Kohn-Sham propagator is typically used
\cite{Hybertsen1985,Godby1986,Northrup1987,Surh1988}. A comparison with the
exact expression (\ref{Eq:self-energy}) identifies two types of omissions:
lack of self-consistency and neglect of vertex corrections. The first group
comprises all terms that allow the self-energy to be written in the
mathematical form (\ref{Eq:GW}) and the polarizability in the form
(\ref{Eq:RPA}) with suitably defined effective propagators. Vertex
corrections, on the other hand, stem from an expansion of the vertex function
$\Gamma$. In this case the topology of the resulting diagrams forbids a
reduction to the simple forms above. Some self-energy diagrams are shown in
Fig.\ \ref{Fig:diagrams}. Arrows represent noninteracting Green functions, the
screened interaction is indicated by a wiggly line. The first diagram
represents the $GW$ approximation. The second is a self-consistency term,
because it can be absorbed into the first by an appropriate renormalization of
the Green function. This is not possible for the third diagram, however, which
therefore represents a vertex correction.

Suitable indicators for the assessment of corrections beyond the $GW$
approximation are their influence on the band structure, the spectral function
and the total energy. The band structure is obtained from
\begin{equation}
E_{n{\bf k}} = \epsilon_{n{\bf k}} + \langle \varphi_{n{\bf k}} |
\Sigma(E_{n{\bf k}}) | \varphi_{n{\bf k}} \rangle \;.
\end{equation}
If the single-particle potential $V_s$ is chosen to include the
exchange-correlation potential of density-functional theory, as is typically
done in practical calculations, then the self-energy must be replaced by the
difference $\Sigma(E_{n{\bf k}}) - V_{\rm xc}$. The quasiparticle energies
feature as prominent peaks in the spectral function, which is proportional to
the imaginary part of the Green function. However, the spectral function also
contains other resonances related to many-body excitations. For instance, a
quasiparticle excitation may be accompanied by a series of satellites that
correspond to the creation or destruction of additional plasmons. Finally,
the total energy is given by the Galitskii-Migdal formula \cite{Galitskii1958}
\begin{equation}\label{Eq:Galitskii-Migdal}
E = \frac{1}{\pi} \int_{-\infty}^\mu d\omega \int d^3r\, \lim_{{\bf r}' \to
  {\bf r}} [ \omega + h({\bf r}) ] \,\mbox{Im}\, G({\bf r},{\bf r}';\omega)
  \;,
\end{equation}
where the one-body Hamiltonian $h$ contains the kinetic energy operator and
the external potential.

\section{Self-consistency}\label{Sec:self-consistency}

Although Hedin's equations define the self-energy as a functional of the Green
function, which is in turn self-consistently determined by the self-energy
through Dyson's equations, most practical calculations simply evaluate the
expression (\ref{Eq:GW}) using the Green function $G_0$ of an appropriate
noninteracting system. This approach is consistent with Hedin's derivation of
the $GW$ approximation and generally yields quasiparticle band structures in
good agreement with experiments. Nevertheless, from a fundamental point of
view this procedure has several shortcomings. First, the use of $G_0$ in the
self-energy introduces an ambiguity, because the result of a $GW$ calculation
then depends on the choice of the single-particle potential $V_s$. Aulbur,
St\"adele and G\"orling \cite{Aulbur2000b} recently investigated this effect
by comparing a $GW$ calculation based on the standard local-density
approximation with one starting from an exact exchange and local-density
correlation potential. Although the substantial difference in the initial
Kohn-Sham band gap is largely levelled out and reduced to about 0.1 eV for
most materials considered, in some cases important quantitative discrepancies
remain. For instance, for GaAs unreconciled band gaps of 1.16 eV and 1.90 eV
are obtained in this way, compared to the experimental value 1.52 eV. Second,
without self-consistency the $GW$ approximation fails to conserve the particle
number, energy and momentum under time-dependent external perturbations
\cite{Baym1961}. Even in equilibrium the integral over the spectral function
\begin{equation}
N = \frac{2}{\pi} \int_{-\infty}^\mu d\omega \int d^3r\, \mbox{Im}\, G({\bf
  r},{\bf r};\omega)
\end{equation}
does not equal the number of electrons \cite{Schindlmayr1997}, as it should,
although the quantitative error is less than 1\% for typical semiconductors
\cite{Rieger1998} and the homogeneous electron gas in the range of metallic
densities \cite{Garcia-Gonzalez2001}. Furthermore, different methods of
calculating the total energy from the Green function are not mutually
consistent \cite{Holm1998}.

Due to the high computational cost, fully self-consistent $GW$
implementations, in which the Green function obtained from Dyson's equation
(\ref{Eq:Dyson}) is used to update the self-energy until convergence is
reached, became possible only a few years ago. The first calculation was
probably performed by de Groot, Bobbert and van Haeringen \cite{deGroot1995}
for a quasi-one-dimensional semiconductor, in which the crystal lattice is
modelled by a sinusoidal potential. The most important finding is a
substantial increase in the band gap, which overestimates the exact
Monte-Carlo result and comes close to the Hartree-Fock gap. Furthermore, the
weight of the incoherent background in the spectral function is drastically
reduced, leading to sharper and more pronounced quasiparticle resonances.

Although the physical characteristics of the model raised doubts concerning
its relevance for real materials \cite{Aryasetiawan1998}, recent calculations
for silicon within a finite-temperature approach have confirmed these findings
\cite{Schone1998}: while the standard non-self-consistent $GW$ approximation
widens the indirect band gap from 0.56 eV in the local-density approximation
to 1.34 eV, in good agreement with the experimental value 1.17 eV,
self-consistency increases the gap to 1.91 eV. The self-consistent $GW$ result
hence overestimates the experimental band gap by as much as the local-density
approximation underestimates it. Besides, self-consistency again leads to an
accumulation of spectral weight in the quasiparticle peaks, in disagreement
with experiments. The quasiparticle peaks are also narrowed, corresponding to
increased lifetimes.

The majority of applications have so far focussed on the homogeneous electron
gas, taking advantage of mathematical simplifications due to the spatial
isotropy. In addition to fully self-consistent results
\cite{Garcia-Gonzalez2001,Holm1998,Eguiluz1998,Holm1999}, several studies have
reported partially self-consistent calculations in which the Green function
$G$ in Eq.\ (\ref{Eq:GW}) is updated until convergence but the screened
Coulomb interaction $W$ is not \cite{Shirley1996,vonBarth1996,Holm1997}. In
contrast to conventional $GW$ calculations, which give an accurate account of
the correlation-induced band narrowing, self-consistency causes the occupied
band width to increase even above its free-electron value $k_{\rm F}^2/2$,
where $k_{\rm F}$ denotes the Fermi wave vector. Incidentially, an
expansion of the valence band width was also found for silicon
\cite{Schone1998}. The weight of the quasiparticle peaks is again increased
and that of the plasmon satellite reduced accordingly. While the quasiparticle
peaks are narrowed, the satellite is broadened and shifted towards the Fermi
level. A calculation for potassium shows that the results of the homogeneous
electron gas can be fully generalized to metals \cite{Schone1998}: the
Kohn-Sham band width of 2.21 eV is narrowed to 2.04 eV in the first-order $GW$
approximation, improving agreement with the experimental value 1.60 eV. The
remaining discrepancy can actually be explained in terms of measurement
effects, which shift the apparent peak position of the Fermi level more
towards lower binding energies than the state at the bottom of the band and
thus give rise to an additional artificial narrowing between 0.2 eV and 0.4 eV
\cite{Shung1988}. In comparison, the self-consistent $GW$ band width is 2.64
eV, much larger even than the Kohn-Sham value.

Taken together, the above results form a consistent picture of the effects of
self-consistency, which in a dramatic way reverses the correct trends of
conventional $GW$ band-structure calculations and destroys the originally good
agreement with experiments. This can be understood as follows. In contrast to
$G_0$, which refers to a noninteracting system with single-particle
excitations only, part of the spectral weight in the interacting Green
function $G$ is transferred to plasmon satellites, which describe collective
excitations. The quasiparticle peaks are reduced accordingly. This
redistribution of spectral weight in turn implies a smaller dynamic
self-energy, i.e., the part $\Sigma_{\rm c} = i G (W - v)$ of the self-energy
that is due to correlation, in the vicinity of the quasiparticle position. The
dynamic self-energy is positive, tending to reduce the band width, and
competes with the exchange part $\Sigma_{\rm x}$, which is always negative and
increases the band width, e.g., by an amount $k_{\rm F} / \pi$ for the
homogeneous electron gas in the Hartree-Fock theory \cite{Mahan1990}. In a
self-consistent calculation the smaller dynamic part no longer dominates over
the exchange part, which is only marginally reduced, and the band width hence
grows rather than narrows \cite{vonBarth1996}. The reduced dynamic self-energy
also explains the larger renormalization factors
\begin{equation}
Z_k = \left( 1 - \left. \frac{\partial \,\mbox{Re}\,
      \Sigma_{\rm c}(k,\omega)}{\partial \omega} \right|_{\omega = \epsilon_k}
      \right)^{-1} \;,
\end{equation}
which indicate the weight of the quasiparticle resonances, as well as the
increase in their lifetime, which is inversely proportional to the imaginary
part of $\Sigma_{\rm c}$. The effects described above are further reinforced
if the screened interaction is also calculated self-consistently
\cite{Holm1998}. In this case the sharp plasmon excitations in $W$ disappear
and it no longer has a physical meaning as a response function.

In conclusion, self-consistency is not a good idea for calculating
quasiparticle energies without the simultaneous inclusion of vertex
corrections. However, an interesting and surprising outcome of the
self-consistent $GW$ calculations for the homogeneous electron gas was the
realization that the total energy obtained from the Galitskii-Migdal formula
(\ref{Eq:Galitskii-Migdal}) is strikingly close to supposedly exact
Monte-Carlo data \cite{Holm1998}. It has been speculated
\cite{Aryasetiawan1998} that this quite unexpected result is related to the
fact that the self-consistent $GW$ approximation is conserving in the sense of
Baym and Kadanoff \cite{Baym1961}, although relative energy conservation under
time-dependent external perturbations does not necessarily imply an accurate
total energy on an absolute scale.

Based on the investigation of finite Hubbard clusters, Schindlmayr, Pollehn
and Godby \cite{Schindlmayr1998a} note that self-consistency systematically
raises the total energy due to an absolute shift of the chemical potential
towards higher energies. Likewise, for the homogeneous electron gas the upward
transfer of spectral weight from low-lying plasmon satellites to the
quasiparticle peak and the increase in the band width, which moves the
quasiparticles to lower energies relative to the chemical potential, work in
different directions and largely cancel. On the other hand, the chemical
potential is shifted upward on an absolute scale, explaining most of the
change in the total energy. Although the increase in total energy often leads
to better agreement with exact numerical solutions for the finite Hubbard
clusters, in some parameter ranges the results become worse. In combination
with the unphysical features of the spectral function described above, these
observations suggest that the good quantitative agreement with Monte-Carlo
data for the homogeneous electron gas may be fortutious. Recent calculations
for the spin-polarized and the two-dimensional electron gas indeed show
slightly larger errors \cite{Garcia-Gonzalez2001}. Nevertheless, these results
have inspired renewed interest in total-energy calculations within many-body
perturbation theory \cite{Holm1999,Holm2000,Sanchez-Friera2000}.

\section{Vertex corrections}\label{Sec:vertex}

Vertex corrections introduce additional interaction channels not accounted for
in the $GW$ approximation (\ref{Eq:GW}) for the self-energy and the
random-phase approximation (\ref{Eq:RPA}) for the polarizability. Their effect
can be understood by physical interpretation. The random-phase approximation
describes dynamic screening within a time-dependent Hartree approach. The
screening electrons around a photoemission hole are thus considered
independent, exchange and correlation, which enforce spatial separation, are
ignored. As a result, the negative charge cloud is too tightly drawn around
the central hole and screening at small distances is overestimated, so much,
in fact, that the pair distribution function in the random-phase approximation
even becomes negative, which is unphysical \cite{Pines1963}. Vertex
corrections in the polarizability will therefore, in general, reduce the
screening and strengthen the interaction. They can also introduce new physical
phenomena, such as bound states between the electrons and holes created during
a photoemission process. In this case the response function acquires an
additional exciton resonance that lies below the plasmon energy. Vertex
corrections in the self-energy describe the same exchange and correlation
mechanisms between the central photoemission hole and the surrounding
particles. The two effects compete and cancel partially: while vertex
corrections in the self-energy reduce the probability of finding other holes
near the central photoemission hole, vertex corrections in the polarizability
simultaneously reduce the screening, thereby increasing the hole density. The
change in the quasiparticle energies will hence be small, albeit important. On
the other hand, the effect on the satellite spectrum may be drastic, because
new types of excitations come into play.

To be consistent, the same vertex function should be used in the self-energy
and the polarizability \cite{Shung1988,Ting1975,Mahan1989}. However, as the
cost of calculating Feynman diagrams grows very rapidly with the topological
complexity, nondiagrammatic vertices or plasmon-pole models for the screened
interaction have long been the only way to determine higher self-energy
terms. Ummels, Bobbert and van Haeringen \cite{Ummels1998} evaluated the
lowest-order vertex correction displayed in Fig.\ \ref{Fig:diagrams} with a
plasmon-pole model for silicon and diamond, revising an earlier calculation
\cite{Bobbert1994}. While the $GW$ approximation increases the direct gap at
the $\Gamma$-point by 0.78 eV for silicon and 2.12 eV for diamond relative to
the local-density approximation, in good agreement with experiments, the
vertex diagram yields an additional contribution of $-0.26$ eV and $-0.09$ eV,
respectively. The results for other high-symmetry points in the Brillouin zone
are similar. It hence appears that the vertex corrections can be numerically
significant, and the reduction in the band gap tends to cancel the increase
due to self-consistency \cite{Schone1998}. However, the use of a plasmon-pole
model leaves some uncertainties, and it is also known from an early
application to the homogeneous electron gas that the lowest-order vertex
correction leads to unphysical analytic properties in the self-energy, which
can produce regions with a negative density of states \cite{Minnhagen1974}.
These unphysical features are only cancelled by higher-order terms.

When vertex corrections are taken into account both in the self-energy and the
polarizability, their effects tend to cancel to a large degree. For the
homogeneous electron gas, Mahan and Sernelius \cite{Mahan1989} calculated the
band width in the range of metallic densities with a static vertex function
and obtained essentially the same result as in the $GW$ approximation,
although there is a strong additional narrowing if vertex corrections are
included only in the polarizability. The latter point was actually noted
before in relation to the band structure of the alkali metals
\cite{Northrup1987,Surh1988}. Likewise, with a vertex function $\Gamma(1,2;3)
= \delta(1,2) f_{\rm xc}(1,3)$, where $f_{\rm xc}(1,3) = \delta V_{\rm xc}(1)
/ \delta n(3)$ denotes the exchange-correlation kernel in the local-density
approximation, Del Sole, Reining and Godby \cite{DelSole1994} found no change
in the band gap of silicon, although the absolute position of the bands is
shifted by about 0.4 eV. This vertex function is obtained by a consistent
first-order solution of Hedin's equations starting from density-functional
theory \cite{Hybertsen1985} as opposed to Hartree theory, which produces the
$GW$ approximation. Stronger changes occur if the vertex function is included
only in the polarizability and not balanced by corresponding self-energy
diagrams: direct gaps are reduced by up to 0.2 eV and the valence band width
decreases by 0.55 eV. Similar conclusions were reached for a finite Hubbard
cluster \cite{Verdozzi1994}.

As expected from the generally good results obtained in the standard $GW$
approximation, vertex corrections, if applied in a consistent manner, tend to
cancel, inducing subtle but quantitatively small changes in the band structure
of solids. Many authors have also reported cancellation between combinations
of vertex and self-consistency diagrams for a variety of systems
\cite{Shirley1996,Ummels1998,deGroot1996}. This observation is reassuring,
because it strengthens the theoretical foundation of the $GW$
approximation. On the other hand, it leaves the question unanswered how
improvements might be achieved. Schindlmayr and Godby \cite{Schindlmayr1998b}
proposed a systematic approach based on a continued iterative solution of
Hedin's equations. As the $GW$ approximation obtained after the first
iteration represents a significant and systematic improvement over the
zeroth-order Hartree or local-density-approximation, further advances might be
achieved by a continuation of this procedure. The implicit integral equation
(\ref{Eq:vertex}) for the vertex function can in fact be solved, and the
functional derivative with respect to the full Green function $G$ is at the
same time replaced by a derivative with respect to the known propagator $G_0$,
which can, in principle, be solved at all levels of iteration. The vertex
correction obtained at the end of the second iteration starting from Hartree
theory mixes diagrams of different order in the screened
interaction. Numerical results for a finite Hubbard cluster show signs of
convergence in the excitation energies and suggest that an iterative solution
of Hedin's equations may improve the spectrum. However, this approach shares
the problem of possible incorrect analytic properties that was earlier noted
for the diagrammatic expansion of the self-energy by orders of the screened
interaction \cite{Minnhagen1974}. Takada \cite{Takada1995a} proposed a similar
approach in which Dyson's equation is solved self-consistently in each
iteration, but so far it has only been exploited to derive a solution method
for model systems without energy dispersion \cite{Takada1995b}.

Incidentially, more progress has been made with respect to systematic
improvements of the satellite spectrum, which from the start is not well
rendered in the $GW$ approximation. In particular, for the homogeneous
electron gas and the alkali metals, the $GW$ approximation only produces a
single plasmon resonance instead of a sequence of satellites separated from
the main peak by multiples of the plasmon frequency. The so-called cumulant
expansion remedies this problem by describing the coupling of a quasiparticle
to multiple plasmons \cite{Langreth1970}. It includes the vertex diagram
displayed in Fig.\ \ref{Fig:diagrams} as well as corresponding higher-order
terms. An application to Na and Al significantly improves the agreement with
experimental photoemission spectra, although the relative intensities of the
satellites with respect to the quasiparticle peak are still in discrepancy
\cite{Aryasetiawan1996}. The quasiparticle position is the same as in the $GW$
approximation. In this context it is interesting to note that with vertex
corrections, the self-energy is less affected by the deterioration of spectral
features, such as the accumulation of spectral weight in the quasiparticle
peak and the broadening of plasmon satellites, when applied self-consistently
\cite{Holm1997}. This is another indicator that the vertex corrections contain
the correct physical features. If the exact vertex function is used, then
self-consistency must, of course, yield the true spectrum without the
mentioned adverse effects.

In a similar way, excitonic effects in the optical absorption spectrum of
semiconductors can be accounted for by vertex corrections that describe
multiple scattering or binding between electron-hole pairs. This procedure is
very computationally demanding, because it requires a calculation of the
two-particle Green function. Nevertheless, excitonic vertex corrections have
been calculated for several semiconductors and significantly improve the
agreement with experimental spectra
\cite{Albrecht1997,Benedict1998,Rohlfing1998}.

\section{Summary}\label{Sec:summary}

The $GW$ approximation is a reliable method for {\em ab initio\/}
electronic-structure calculations that produces band structure in good
agreement with experiments for a wide range of materials. This was originally
taken as a sign that higher-order self-energy diagrams are negligible.
Statements to this effect indeed frequently appear in the early literature
\cite{Hedin1969}. Only the availability of modern computational resources in
recent years has it made possible to explicitly evaluate self-consistency and
vertex corrections and test this assumption. The numerical results show that
the excluded terms are, in general, not individually small but tend to
mutually cancel. The findings of self-consistent calculations are consistent
and show a serious deterioration of spectral features compared to the standard
$GW$ approximation, which can be understood by shifts of spectral weight in
the dynamic self-energy. The influence of vertex corrections is naturally less
clear-cut, reflecting the large variety of possible vertex functions. However,
the case for a mutual cancellation of vertex corrections in the self-energy
and the polarizability is well established. Although a general and systematic
way of improving quasiparticle energies is still outstanding, physically
motivated vertex corrections for better satellite spectra are known.

\section*{Acknowledgments}

This work was funded in part by the EU through the NANOPHASE Research Training
Network (Contract No.\ HPRN-CT-2000-00167).

\end{document}